\documentclass[a4paper,graphics,natbib,graphicx,psfig,amssymb,ccaption,layout]{article}
\usepackage{lscape, morefloats , graphicx, float}
\newcommand{\affil}[1]{$^{\rm #1}$}
\usepackage[authoryear]{natbib}
\bibpunct{(}{)}{;}{a}{}{,}
\date{} %Please leave the date blank

\title{\large\bf\flushleft The Catalogue of Stellar Parameters from the Detached Double-Lined Eclipsing Binaries in the Milky Way}
\author{\parbox{\textwidth}{\flushleft
\vspace{-0.5cm}
{\it Z. Eker\affil{A, *}, S. Bilir\affil{B}, F. Soydugan\affil{C, D}, E. Yaz G\"ok\c ce\affil{B}, E. Soydugan\affil{C, D}, M. T\"uys\"uz\affil{C, D}, ~~~~T. \c Seny\"uz\affil{C, D}, O. Demircan\affil{D,E}}\\
\vspace{0.4cm}
{\small \affil{A}\,Akdeniz University, Faculty of Sciences, Department of Space Sciences and Technologies, 07058, Antalya, Turkey}\\
{\small \affil{B}\,Istanbul University, Faculty of Sciences, Department of Astronomy and Space Sciences, 34119, University-Istanbul, Turkey}\\
{\small \affil{C}\,\c Canakkale Onsekiz Mart University, Faculty of Sciences and Arts, Department of Physics, 17100 \c Canakkale, Turkey}\\
{\small \affil{D}\,\c Canakkale Onsekiz Mart University, Astrophysics Research Center and Ulup\i nar Observatory, 17100 \c Canakkale, Turkey}\\
{\small \affil{E}\,\c Canakkale Onsekiz Mart University, Faculty of Sciences and Arts, Department of Space Science and Technologies,\\ 17020 \c Canakkale, Turkey}\\
{\small \affil{*}\,Email: eker@akdeniz.edu.tr}}}
\begin{document}
\twocolumn[
\begin{changemargin}{.8cm}{.5cm}
\begin{minipage}{.9\textwidth}
\vspace{-1cm}
\maketitle
\small{\bf Abstract:}
The most accurate stellar astrophysical parameters were collected from the 
solutions of the light and the radial velocity curves of 257 detached double-lined 
eclipsing binaries in the Milky Way. The catalogue contains masses, radii, 
surface gravities, effective temperatures, luminosities, projected rotational 
velocities of the component stars and the orbital parameters. The number of 
stars with accurate parameters increased 67 per cent in comparison to the most 
recent similar collection by \citet{Torres10}. Distributions of 
some basic parameters were investigated. The ranges of effective temperatures, 
masses and radii are $2750<T_{eff}$(K)$<43000$, $0.18<M/M_{\odot}<33$ and 
$0.2<R/R_{\odot}<21.2$, respectively. Being mostly located in one kpc in the 
Solar neighborhood, the present sample covers distances up to 4.6 kpc within 
the two local Galactic arms Carina-Sagittarius and Orion Spur. The number of 
stars with both mass and radius measurements better than 1 per cent uncertainty 
is 93, better than 3 per cent uncertainty is 311, and better than 5 per cent 
uncertainty is 388. It is estimated from the Roche lobe filling factors that 
455 stars (88.5 per cent of the sample) are spherical within 1 per cent of 
uncertainty.

%%%%%%%%%%%%%     KEYWORDS    %%%%%%%%%%%%%
\medskip{\bf Keywords:} Stars: fundamental parameters -- Stars: binaries: eclipsing 
-- Stars: binaries: spectroscopic -- Astronomical Data based: catalogues
% Please write all keywords in lower case. PASA uses the
% standard list of subject headings adopted by The Astrophysical Journal
% and available from http://www.journals.uchicago.edu/ApJ/keywords_text.html.
%Keywords are separated by em-dashes, i.e. ---
%%%%%%%%DO NOT EDIT%%%%%%%%%%%%
\medskip
\medskip
\end{minipage}
\end{changemargin}
]
\small
%%%%%%%%EDIT FROM HERE%%%%%%%%%%%%

\section{Introduction}
Nearly sixty percent or more of the Solar neighborhood stars are binaries or 
multiple systems \citep{Duquennoy93}. Binaries are important for 
astrophysicists not only because they over populate single stars, but also 
because they provide basic stellar parameters as independent observed 
quantities used in testing astrophysical theories. Stellar masses can be 
determined directly via application of Kepler's law for the visual binaries 
if apparent orbital parameters were calibrated to be real. A calibration is 
possible for a visual pair if its distance (parallax) is known. Reliable 
stellar masses could also be obtained from the radial velocity curves 
without distance, but only if orbital inclinations were known. The resolved 
double stars (visual binaries confirmed to be spectroscopic binaries), 
therefore, are a special case to provide reliable stellar masses, since 
the orbital inclinations are from the apparent orbit and the absolute 
orbital sizes are from the radial velocity curves. 

Gainfully, the light curves of eclipsing binaries could provide orbital 
inclinations and the radii relative to the semi-major axis of the orbit. 
But, if both stars are resolved spectroscopically, accurately determined 
radii and masses could be obtained from the simultaneous solutions of the 
light and the radial velocity curves. In addition to the radii, which 
are not available from visual binaries, the eclipsing spectroscopic 
binaries provide masses, effective temperatures and the absolute dimensions 
of the orbit, from which absolute brightness's could be calculated. 
Provided with a parallax, either the physical parameters or the parallax 
could mutually be tested by comparing the photometric and the trigonometric 
distance moduli. Otherwise, a proper solution would provide not only the 
most reliable stellar parameters, but also a reliable photometric distance 
(parallax) as an independent quantity.

The critical compilations of stellar parameters and absolute dimensions of binary 
components were initiated, and continued with increasing quantity and quality 
especially by \citet{Popper80} and \citet{Harmanec88}. \citet{Andersen91} collected 
accurate stellar masses and radii with uncertainties less than 2 per 
cent from the detached, double-lined eclipsing systems. The list contained 45 
(90 stars) binaries which are all non-interacting, so that each star could be 
accepted as if evolved as single stars. Accuracies of 1-2 per cent were 
found to be significant for deeper astrophysical insight than merely improving 
the spectrum of masses and radii. Due to great sensitivity of other parameters, 
only limited amount of useful results could be extracted up to $\pm$5 per cent 
uncertainties.

\citet{Malkov93} announced a catalog of astrophysical parameters of binary systems 
containing 114 systems including all pre and out of main sequence, contact and 
semi-contact systems. \citet{Gorda98} collected stellar masses and radii with 
accuracies better than 2-3 per cent from photometric, geometric, and absolute 
elements of 112 eclipsing binaries with both components on the main sequence, 
for studying the mass-luminosity and mass-radius relations. Ages and 
metallicities for the components of 43 eclipsing binaries with lines of both 
components visible on the spectra were studied by \citet{Kovaleva01}.

As the number of stars with reliable physical parameters is increasing, the
studies concentrated more on the precision and accuracy. Therefore, \citet{Ibanoglu06} 
did not combine 74 double-lined detached eclipsing Algols and 61 semi-detached 
Algols when plotting mass-radius, mass-T$_{eff}$, radius-T$_{eff}$, 
and mass-luminosity data. \citet{Lastennet02} compiled 60 non-interacting,
well-detached systems with typical errors smaller than 2 per cent for masses
and radii, while 5 per cent for the effective temperatures. The core
of the sample was the large catalogue of \citet{Andersen91}. As being satisfied 
with 10 per cent accuracy for the main-sequence stars, \citet{Hillenbrand04} 
studied dynamical mass constraints on pre-main-sequence evolutionary tracks with 
148 stars, 88 are on the main-sequence, 27 are on the pre main-sequence and 33 are 
on the post main-sequence, where the source of data were \citet{Andersen91}, 
\citet{Ribas00} and \citet{Delfosse00}. Despite, the number of eclipsing 
binaries was 6330 in ``the catalogue of eclipsing binaries'' by 
\citet{Malkov06}, but \citet{Malkov07} was able to select only 215 stars 
(114 binaries) which are detached-main-sequence and double-lined 
eclipsing binary with uncertainties for masses, radii, $\log T_{eff}$, 
$\log L$, and $\log M$ were assumed to be 10, 10 per cent, 0.03, 0.03 
and 0.1 mag or better respectively.

Eclipsing binaries are not only recognized with their accuracy, but also 
known to have larger spectrum of mass range especially towards larger 
masses in compared to visual binaries including {\em Hipparcos} detections  
\citep{Malkov03}. Improvements in observing and analysis techniques never 
stop, and collection of reliable light and radial velocity curve solutions 
continue. Recently, \citet*{Torres10} updated the critical compilation of 
detached, double-lined eclipsing binaries with accurate masses and radii. 
Superseding \citet{Andersen91} list, this new list contains 190 stars (94 
eclipsing binaries and $\alpha$ Cen). In order to fill the gaps in different 
mass ranges, further compilations on the accurate absolute dimensions of the 
eclipsing double lined spectroscopic binaries are inevitable. 

The aim of this paper is to present our compilation of 514 stars which are 
from 257 detached eclipsing double-lined spectroscopic binaries (SB2) of 
the Milky Way. The number of stars in our list with reliable masses, 
spectroscopic mass ratios, orbital inclinations, radii, $T_{eff}$, $\log g$, 
and $v\sin i$ values as collected from the literature supersede similar 
compilations before. Selecting criteria and data collection from binaries and 
the descriptions of the quality of the data are given in Section 2. The H-R diagram, 
space distributions and how good a typical single star represented were 
discussed in Section 3 and finally conclusions were provided in Section 4.   

\section{The Catalogue}
\subsection{Selecting Criteria and Data Collection}
The basic stellar (masses, radii, $T_{eff}$, $\log g$, $v\sin i$)
and orbital (radial velocity amplitudes $K_1$, $K_2$, mass ratio $q$, orbital
period $P_{orb}$, orbital inclination $i$, semi-major axis $a$, eccentricity $e$)
parameters have been collected from the published material through literature  
from the simultaneous solutions of light and radial velocity curves of 257 
detached eclipsing and double-lined spectroscopic binaries in our Galaxy. The 
collections are complete by the date January 2, 2013. The strict initial 
criterion was first to make an eye inspection to the light 
and the radial velocity curves, to make sure the system is detached, and have 
sufficient number of observed data on both curves to assure acceptable accuracy. 
The systems having W UMa and $\beta$ Lyr type light curves, which are known as 
contact and semi-contact systems respectively, were avoided. The near contact 
systems showing noticeable proximity effects such as ellipticity and reflection 
on their light curves were also avoided. Another criterion was to make sure all 
systems are in the Milky Way, so to avoid complexities involving extragalactic 
origins and to form a homogeneous sample.

If there were such systems studied more than once, that is, the light and the 
radial velocity curves determined and solved in various studies, the most recent 
solutions which include or compare to the previous solutions were preferred. 
We were aware of the fact that some of the chromosphericaly active binaries are 
possible to fulfill the above criteria. For example, if a detached, double-lined 
eclipsing binary is one of the studied chromosphericaly active binaries with 
starspots, we were keen to accept stellar parameters only if spotless clean 
solutions exist, in order to avoid negative contribution of starspots on the 
stellar parameters. Therefore, 24 binaries in the current collected sample are 
also contained in third edition of the catalogue of Chromosphericaly Active 
Binaries (CABs) by \citet{Eker08}.

\subsection{Impact of Selecting Criteria and Observational Bias}
Binaries have numerous advantages over single stars when determining basic 
stellar parameters, like masses, radii, effective temperatures, etc, through 
observations. Direct measurements of stellar radii come only from the light 
curve analysis of eclipsing binaries. Therefore, the first selection criterion 
is to require eclipses.

After many years of observational experience today for recognizing 
eclipsing nature of a light curve, it's always useful for the one to 
remember the famous debate, binary versus pulsation \citep{Shapley1914, 
Eddington1918, Hall94}, in the beginning of the twentieth century. Even if 
the one is sure, the light curve is in eclipsing nature, the light curve 
alone is not sufficient to extract absolute sizes of the orbit and radii of 
the component stars unless apparent angular size and distance of the eclipsing 
system is known independently. Radial velocity curves of eclipsing SB2 
binaries not only could confirm consistency of eclipses but also provide 
the absolute size of the orbit by using the value of orbital inclination 
from the light curve analysis. Consequently, absolute sizes of radii could be 
computed from the eclipse durations and absolute size of the orbit. Therefore, 
for the second selection criterion, each eclipsing binary must be an SB2 system. 

Eclipsing light curves and radial velocity curves of SB2, however, does not 
always assure basic stellar parameters (mass, radius, effective temperature, 
surface gravity, luminosity, etc.) as accurate as usable by the theoretical 
astronomers to compare and test their astrophysical theories. Therefore, complex 
cases associated with these two basic criteria must be removed by additional 
selection rules. So, the third selection criterion is established as ``avoiding 
extragalactic systems'' in order to form a homogeneous sample representing 
binaries in the Milky Way and in the Solar neighborhood.

Most stellar astrophysical theories start with the simplest assumptions, e.g. 
being single, spherical, non-rotating, not-pulsating etc. Complexities, such 
as mass loss, convective overshoot, rotation, pulsation etc were added later.
Today there are many theoretical stellar evolution models \citep[i.e.][]{Pols98, 
Demarque04,Girardi10, Ekstrom12} competing. Apparently those models, which 
involve single stars, were tested and/or improved by comparing observational 
parameters mostly comes from binaries advantageous to provide most accurate 
data. Moreover, single star evolution becomes invalid if it is an interactive 
binary and experience mass transfer. Therefore, similar kinds of data from 
single stars are not only poor thus not useful but also not free from many 
severe complexities. For example, one can never be sure a single star is 
really single and evolved as a single. \citet*{demink11} claimed ``the only 
unambiguous identification of true single stars is possible in detached 
binaries, which contain two-main sequence stars''. Therefore, for the fourth 
selection criterion in this study, ``avoid interacting binaries known to be 
involving mass transfer, such as W UMa and $\beta$ Lyr types''.

Near contact binaries, with strong proximity effects, still introduce 
complexities involving strong tidal synchronization, deformation of shapes 
and mutual irradiations of the component stars. So, fifth selection criterion, 
``avoid systems with proximity effects'', must be added. 

Active solar-type stars with starspots may introduce minor but unavoidable 
complexity causing determined radii, relative temperatures to be less accurate. 
Therefore, sixth selection criterion, ``avoid stellar activity, accept the 
spotless solutions''.

\subsection{Description and Quality of the Data}
The catalogue is available in an electronic format (Table 1). The content 
of the catalogue are organized as a table of 257 rows and 60 columns. Thus, 
each row of data belongs to a detached eclipsing binary, which is an SB2 
system. The columns and their descriptions were given in Table 2, where one 
may notice the data were carefully referenced for interested readers who 
want to see original sources. Original sources, however, are heterogeneous 
to use the older and the newer values of $GM_{\odot}$ and $R_{\odot}$ when 
deriving the observed masses and radii. Since contributing uncertainty of those 
constants is not entirely negligible (0.23 per cent) as estimated by 
\citet{Torres10}, by means of their formulae, the collected masses and radii 
were homogenized using $GM_{\odot}=1.3271244\times10^{20}$ m$^{3}$s$^{-2}$ 
\citep{Standish95} and $R_{\odot}=6.9566\times$10$^{8}$m \citep*{Haberreiter08}. 
Next, surface gravities and luminosities recomputed from the homogenized 
quantities. Those re-evaluated quantities and their associated errors are 
listed in Table 3 where columns are self explanatory.    

Containing 514 stars (257 binaries), the present catalogue supersedes the 
most recent collection by \citet{Torres10} with 190 stars (94 eclipsing 
systems and $\alpha$ Cen). Both catalogues are similar in a sense 
containing detached eclipsing binaries with an exception that $\alpha$ 
Cen being astrometric binary is excluded. Moreover, extragalactic binary 
OGLE 051019, which is in the list of \citet{Torres10}, is also excluded 
because it is not in our Galaxy. As a result, only 93 binaries are common in 
both catalogues, hence there are substantial amount of excess 
(257-93=164) in the present catalogue. However, it is prudent to 
compare the quality of the data rather than a plain number. \citet{Torres10} 
preferred to collect binaries with the masses and radii of both stars to be 
known within errors of $\pm3$ per cent accuracy or better. Rather than 
binaries, present study concentrates on component stars 
and collects individual stars with most accurate masses and radii. In the 
present sample, the number of stars both mass and radius within $\pm3$ per 
cent accuracy or better is 311. The number implies 67 per cent more stars 
than \citet{Torres10} with similar accuracy. Among the 311, the 292 stars 
are matched binaries,  thus the number of Galactic binaries with similar 
accuracy increased from 93 to 146 (57 per cent) in this study. When 
comparing the most accurate data, that is, the number of individual stars 
both mass and radius with accuracies $\pm1$ per cent or better are 93 and 
43 in the present list and in the list of \citet{Torres10}, respectively. 
Such improvements justify the publication of the present catalogue. 

Among the 93 binaries which are common between the present catalogue and 
the catalogue of \citet{Torres10}, only one binary, EW Ori, has been 
re-studied later by \citet{Clausen10}. Thus, 92 binaries in both 
catalogues have common references from which catalogue parameters were taken. 
Therefore, catalogue values and accuracies must be similar. Nevertheless, 
we have noticed few small negligible differences most probably originating 
from the preferences among the multiple references. Additionally, few limited 
non-negligible differences exist because of identifying the primary and the 
secondary, which will be explained below while discussing the mass ratio 
($q$) distribution of the present sample.

Statistics of masses, radii, effective temperatures, surface gravities, 
luminosity of individual stars, and orbital semi-major axis ($a$), mass ratio 
of binaries in the present sample are summarized in Table 4. Maximum and minimum 
values of those data with the identified star or the system are given in the 
first four rows. Following are mean, mode and median values. Maximum, 
minimum, mean, mode and median values of the associated errors are given in the 
last five lines. Relative errors are indicated by \% sign after the value, 
otherwise the errors are absolute. Being related with characteristics of the 
present sample, some of the distributions will be discussed below.  

\subsubsection{Apparent Magnitude and Period Distributions}
Apparent magnitude distribution of the present sample (257 systems) is 
shown in  Fig. 1. Large range of apparent magnitudes from very bright 
(1.89 mag, $\beta$ Aur) to dimmest (19.92 mag, SDSS-MEB-1) display a 
peak at $V=9$ mag, thus most of the systems are contained at 8, 9 or 10 
magnitudes. 

The orbital period of the sample covers a range from $P_{orb}=0.30$ days (DV Psc) 
to $P_{orb}=99.7638$ days (V379 Cep). The distribution on the logarithmic scale 
is displayed in Fig. 2. Accordingly, the most common orbital periods are 
close to 2.5 days. For a given mass, bigger orbital periods mean larger 
orbital sizes. Larger orbital size, however, decreases the probability of 
having eclipses. Since, our sample contains only eclipsing binaries, the 
decrease towards longer periods in Fig. 2 is clear and it must be effected 
by the probability of eclipses. Researchers prefer to study short period 
systems because they are easier to observe. This is an additional bias to 
increase the number of systems towards the short periods. However, according 
to Fig. 2, starting from $\log P_{orb}$ (days)$=0.4$, the number of systems decreases quickly 
towards the short periods. This could be explained by the fact that decreasing 
orbital period means decreasing the orbital sizes. Decreasing orbital size, 
however, increases the proximity effects such as, reflection and deviation 
from spherical shapes. Avoiding systems with proximity effects must be the 
main cause of the decrease towards short periods starting from the peak period 
$P_{orb}=2.5$ days.   

Distribution of orbital periods among the spectral types (primary 
components) is given in Table 5. Shortest orbital periods (0.30 to 0.63 
days) could be found among F and later spectral types. Such small orbital 
periods does not exist for the binaries with spectral types O, B and A. 
The system with the smallest orbital period ($P_{orb}=0.30$ days, DV Psc) has 
components of spectral types K4V+M1V \citep{Zhang07}.

Among the O-type binaries, the shortest orbital period is 1.62 days 
which belongs to V1182 Aql \citep{Mayer05}, while the longest orbital period 
is only 4.24 days for V1292 Sco \citep{Sana06}. It appears odd to find the 
maximum orbital period of O-type systems as short as 4.24 days. The long period cut of 
for the B-type systems is at 99.76 days which is the orbital period of V379 Cep 
\citep{Harmanec07}. Normally, larger radii would increase the probability of 
eclipses, thus, one would expect to see more O-type systems thus the median values 
of the orbital periods (Table 5) must increase towards earliest spectral types. 
However, the median $\log P_{orb}$ decrease from G-type to O-type. Moreover, according 
to Table 5 (column 9), there are 247 binaries from B to M types. Consequently, 8 out 
of 247 with $\log P_{orb}$ (days) $>1.4$, and 32 out of 247 with $\log P_{orb}>1$ hence, 
probability to have chance 0 out of 10, which is the case for O-type systems, is 0.72 
($\log P_{orb}$ (days) $>1.4$) and 0.25 ($\log P_{orb}$ (days) $>1$) 
not compellingly low. These numbers might be biased somehow to very high values, but 
even computing only on the B-type stars (disadvantage of smaller subset), 1 out of 52 
and 8 out of 52 gives probabilities of 0.82 ($\log P_{orb}$ (days) $>1.4$) and 0.19 
($\log P_{orb}$ (days) $>1$) to have no O-type stars at longer periods. The 
apparent lack of O-type stars at longer periods suffers from statistical significance 
because of the small size of this subset. However, inspecting the median values 
of mass ratios (Table 5), one may notice that the median values of mass ratio 
($q$), is almost constant (median $q$=0.92) for late spectral types (G, K, M), 
but it decreases from F-types to O-types as low as median $q=0.708$. This implies 
that, O-types stars have higher probability to have a secondary with less massive 
than the primary. Combining this with faster rotation, the odds of detecting the 
secondary due to rotational broadening increases. Thus, faster rotation of 
O-type stars appears to be the main reason finding them with the least number and 
smallest upper cut off for the orbital periods on the Table 5.   

\subsubsection{Masses and Radii} 

The distributions of masses and radii of 514 stars are shown in Fig. 3. 
Similar appearance of both distributions is not a coincidence. This is 
because, almost all of the stars of the present sample are on the main 
sequence, only a few exceptions e.g. BW Aqr, V379 Cep, RT CrB, TZ For, AI Hya, 
V1292 Sco appear to have evolved components. Similar appearance of the two 
distributions, thus, imply a strong relation between stellar masses and radii, 
that the shape of both distributions is almost identical.

The uncertainty distributions of masses and radii are displayed in 
Fig. 4. Fig 4a, c are in the accumulated form and Fig. 4b, d are ordinary 
histogram showing the number of stars in 1 per cent bin up to 20 percent 
uncertainty. Fig. 4a, c indicate that in the sample there are 193 stars 
with masses better than 1 per cent, 390 stars with masses better than 3 
per cent, and 443 stars with masses better than 5 per cent uncertainties. 
Regarding to radii in the sample, there are 158 stars with radii better 
than 1 per cent, 379 stars with radii better than 3 per cent, and 437 stars 
with radii better than 5 per cent uncertainties. The uncertainty range up to 
20 per cent includes all mass data, while the radii data contain 98 per cent 
of sample stars in the same range. That is, there are only nine stars with 
uncertainties larger than 20 per cent in radius. 

\subsubsection{Mass Ratio and Spectral Types}
The spectroscopic mass ratio ($q$) of the components is one of the most 
fundamental parameter of binary systems. It is conventionally defined as 
mass of the secondary divided by the mass of the primary. If the secondary is 
the less massive component, $q$ is always less than one. Here, the light curve
approaches are adopted; the primary is the eclipsed star during the primary
minimum which is the deeper one of the two minima. Consequently, all mass 
ratios are determined spectroscopically from the radial velocity amplitudes 
as $K_1/K_2$.

The mass ratio distribution of the sample stars is displayed in Fig. 5.
The most common mass ratio of the sample is close to 0.9 and it is
distributed within the range 0.2 to 1.2. The values $q<1$ indicate more
massive hotter primary and a cooler less massive secondary. This is the
case for young unevolved binaries on the main sequence. On the other hand, 
the values $q>1$ implies evolved pairs that more massive component cooler 
than the less massive secondary. Even when there would be no evolved pairs, 
some empirical mass ratios above unity expected because of less precise 
systems with a $q$ near unity. According to Fig. 5, there are 21 systems 
with $q>1$. Examining their positions on the H-R diagram, seven systems 
(TZ For, V1130 Tau, AI Hya, GZ Leo, RT CrB, V2368 Oph, V885 Cyg) were found 
for sure to have evolved cooler components. The rest have some ambiguity 
because one can not identify the primary and the secondary due to 
component masses and/or temperatures so close. 

Using published spectral types (Table 1) and counting each component as a 
single star, spectral type distribution of 514 stars (257 pairs) is shown 
in Fig. 6. Except for O-types, there exist statistically sufficient number 
of stars at all spectral types. The most crowded spectral types are F, A and B. 
So, with an abrupt increase at B, number of stars increase until F, then 
there is a considerable sudden decrease to G then gradual decrease towards 
M. Readers should be aware of that  the papers announcing those spectral 
types (Table 1) use different kinds of estimates, from professional MK(K) 
types to very rough estimates, even from photometric indices.

The hottest star in the sample has an effective temperature 43000K which 
is the O-type primary of V1182 Aql \citep{Mayer05}, while the coldest star 
in the sample has an effective temperature 2750K which is M-type secondary 
of KIC10935310 \citep{Cakirli13}.  

\section{Discussion}
\subsection{H-R diagram}
The H-R diagram is a primary tool to demonstrate and study stellar evolution. 
Representing the stars with most accurate mass and radius, the present sample 
on the H-R diagram has been investigated. Taking the effective temperatures 
($\log T_{eff}$) from the catalogue (Table 1) and using the homogenized 
luminosities ($\log L_{\odot}$) from Table 3, the positions of 472 (236 
binaries) stars on the H-R diagram are shown in Fig. 7. Theoretical stellar 
evolution lines \citep[from][]{Pols98}, ZAMS and TAMS are drawn to 
indicate evolutionary status of the current sample. Almost the entire sample 
appears to be within the main-sequence band. Evidently, there are no supergiants 
despite the selection criteria do not imply such a result directly. Considering 
the fact that evolved stars could be observed at larger distances than their 
main-sequence counterparts and the probability of having eclipses is higher 
if the radii of the components are larger, the absence of giants and 
supergiants in the present sample needs to be explained. Smaller number of 
evolved stars, which is a consequence of shorter lifetime of evolved phases, 
and low probability of evolving both components to the sizes comparable to 
each other, which permits the detection of eclipses, must be more effective 
than the eclipse favoring conditions. Obviously, much smaller size of 
companion stars if exists, and/or huge difference of brightness between the 
components leave out some giants and supergiants to be detected as SB2 
eclipsing binaries.   

Although there are 514 stars (257 binaries) in the present sample, we are able 
to place 472 stars (236 binaries) with effective temperatures on the H-R diagram. 
This is because the catalogue (Table 1) has 21 systems (42 stars have no published 
temperatures; see Table 1) without temperatures. When solving the light and the 
radial velocity curves, some authors, who are not trusting stellar temperature 
determinations, are satisfied with the temperature ratios rather than absolute 
temperatures \citep[i.e.][]{Helminiak09}. The obtained masses and radii for 
those 21 systems are still reliable. Therefore, they were included in the catalogue. 

Sample stars on H-R diagram were displayed as four sub-samples according to accuracy 
limits. The four sub-samples [both mass and radius up to $\pm1$ per cent accuracy 
91 stars, between $\pm1$ to $\pm3$ per cent (211 stars), between $\pm3$ to $\pm5$ 
per cent (71 stars) and worse than $\pm5$ per cent (99 stars)] 
were shown  in Fig. 7a with different symbols in order to investigate if there are 
any preferred positions among the data of different class of accuracy. Except for 
the most accurate sub-sample (91 stars) which covers a spectral range from A0 to M3 
corresponding to a temperature range 10000 K to 3000 K, the other sub samples have 
full ranges covering spectral types from O5 to M3 and temperatures from 43000 K to 
3000 K. Preference of certain locations by certain sub-samples cannot be noticed 
except the most accurate sub-sample. All other sub-samples appear to be evenly 
distributed along the main-sequence band.

In order to investigate accuracy and precision of the positions on the H-R 
diagram (Fig. 7), the published uncertainties of radii and temperatures have 
been propagated to estimate the uncertainty of the luminosities. Among the 
472 stars (236 binaries) plotted on Fig. 7a, the seven systems 
(2MASS J01542930+0053266, 2MASS J16502074+4639013 and 2MASS J19071662+4639532, 
GZ Leo, HY Vir, $\delta$ Vel, V415 Aql) do not have published uncertainties for 
their temperatures and seven systems (V467 Vel, IQ Aqr, SZ Cam, RX Her, V2083 Cyg, 
MR Del, AE For) were found to have uncertainty of temperatures only for their 
secondaries. Leaving the systems without any temperature uncertainty out, and 
assigning same uncertainty to the primary temperature for those having only 
uncertainty for the secondary temperatures, the relative uncertainties of 
luminosities for 458 stars (229 binaries) were computed by the method of error 
propagation. All computed uncertainties are less than 51 per cent except for 
one star, which is the secondary of V1292 Sco. The 18.8 per cent uncertainty 
of the temperature and 51.9 per cent uncertainty on the radius propagate to be 
128 per cent for the luminosity for this star. Among 458 stars, 87.6 per cent 
(401 stars) have relative errors less than 20 per cent. The rest, 57 stars are 
placed on the H-R diagram with their error bars (Fig. 7b) in order to investigate 
if certain locations are preferred by the stars with the biggest error bars. 
Evenly distribution along the main-sequence band covering the full range of 
spectral types is clear. However, one must remember that the most accurate 
sub-sample (91 stars) does not mean to have the most accurate positions on the 
H-R diagram because an uncertainty at an effective temperature contributes at both 
axes. Consequently there are contributions from all subsamples to Fig. 7b. 
Nevertheless, it is interesting to notice the least accurate sub-sample dominates 
(44 per cent) although it occupies only 21 per cent (99 stars) among 472 stars 
in Fig. 7a. The contributions to Fig. 7b from the other three are sub-samples are 
5, 18, and 9, which sum up to be 56 per cent, respectively from the groups of 
91, 211 and 71 stars of the most accurate and the other two subsamples in Fig. 7a.

Determination of observed temperatures appears to be the biggest obstacle of 
observational astrophysics to study stellar evolution on the H-R diagram. 
Considering the fact that, some authors prefer to publish 
internal errors, which could be unrealistically small, e.g. the effective 
temperatures of AE For $T_{eff}$(pri)$ =4100\pm6$ K and $T_{eff}$(sec)$ =4055\pm6$ K 
\citep{Rozyczka13}, XY UMa $T_{eff}$(pri)$ =5200\pm7$ K and $T_{eff}$(sec)$ =4125\pm7$ K 
\citep{Pribulla01}, and DV Psc $T_{eff}$(pri)$ =4450\pm8$ K and $T_{eff}$(sec)$ =3614\pm8$ 
K \citep{Zhang07}, many of the uncertainties on the effective temperatures of the present 
sample are optimistic. But still, the present sample has the most accurate stellar 
positions on the H-R diagram ever been obtained from the simultaneous solutions of 
light and radial velocity curves of detached eclipsing binaries and the stellar 
parameters would be the most reliable, thus, empirical relations and astrophysical 
theories could be tested.

\subsection{Space Distributions}
Distributions of the present sample on the equatorial and the Galactic coordinates 
are displayed in Fig. 8. Although 257 systems appear to be homogeneously distributed, 
a significant asymmetry between northern and southern hemispheres is indicated by the 
numbers in the lower right corner of Fig. 8a. There exist 168 northern binaries with 
positive declinations ($\delta \geq 0^{o}$), while there are nearly half number of 
binaries in the southern hemisphere, that is, 89 systems with negative declination 
($\delta < 0^{o}$). Detailed analyses is not in the scope of the present paper, but 
in the first approximation, the excess in the northern hemisphere could be explained by
overpopulation of the northern telescopes and/or astronomers to the southern ones. 
A perceptible concentration of stars towards the Galactic plane is noticeable even 
on the equatorial coordinates that the Galactic plane drawn on Fig. 8a, where the 
Galactic center is marked by a star symbol.

Although a negligible north-south asymmetry is indicated by the numbers at the 
lower left corner, nearly symmetric distribution with respect to Galactic plane 
was shown on Fig. 8b. It is also noticeable that there are regions on 
the Galactic plane where the stars appear to be grouping towards the Galactic 
longitudes 30, 70 and 110 degrees. These are the directions associated with 
the local arms structure of our Galaxy. On the other hand, less populated, 
``empty region'' towards $l=250$, $b=-45$ degrees are also noticeable on the 
Fig. 8b. 

Among the 257 systems in the present sample, 205 systems were found to
have distances published at articles from where catalogue data were taken. 
Therefore, we had to search extra sources for the distances of 52 systems, 
where 24 of them were found to have {\em Hipparcos} parallaxes 
\citep{vanLeeuwen2007}. For the rest with no trigonometric parallaxes, 
the formulae by \citet{Bilir08a}, was used in estimating distances. Finally, 
the distance distribution of the sample in the Solar neighborhood were 
presented in Fig. 9. The Fig. 9b indicates the number of systems within each 
incremental 200 pc bins. Accumulated form of the same data (Fig. 9a) 
indicates that 90 per cent of sample stars are contained within 1 kpc from 
the Sun, where the rest are located at distances up to 4.6 kpc. Such 
distances are far more than the detection limit of {\em Hipparcos} 
\citep{Perryman97}. In fact, excluding the parallaxes with large relative 
errors (e.g. $\sigma_{\pi}/\pi>0.5$) in the sample, there are 119 systems 
(46.3 per cent) with reliable parallaxes, thus, more than half (53.7 per 
cent) of the sample had to rely on the photometric distances. Compared to 
other field stars, the present sample are advantageous to provide the most 
accurate photometric distances together with physical parameters.

Figure 10 displays space distribution of the sample in the Solar 
neighbourhood on the Galactic plane ($X$-$Y$), where $X$ is towards the 
Galactic center and $Y$ is towards the Galactic rotation. With a median 
distance of 1528 pc, O-type binaries are the most distant objects. 
Closest one is SZ Cam which is 870 pc away \citep{Tamajo12}, the most distant 
one is DH Cep with 2767 pc \citep{Hilditch96}. There exist 10 binaries with O-type
primary in the present sample. The distant stars, mostly O-B type binaries on 
the $X$-$Y$ plane were interesting (Fig. 10a) as if to imply one of the local 
Galactic arms. Indeed, after plotting, O-type binaries in our sample on Fig. 11, 
which shows Galactic arm structure by \citet{Xu09}, the position of 
O-type binaries and the Sun on the Galactic plane became clear. The present 
sample is mostly located within 1 kpc in the Solar neighborhood which is itself 
positioned between the two local arms Carina-Sagittarius and Orion Spur. The Orion 
spur itself situated between the Perseus and Carina-Sagittarius arms.

Figure 10b shows distribution perpendicular to the Galactic plane, where the 
scaleheight of the thin disc ($H=220$ pc) according to \citet{Bilir06a,Bilir06b, 
Bilir08b}. Distant systems, mostly with O-B type binaries on the Galactic plane 
are distinguishable from the concentrated central region.

On a closer look, the space distribution of the concentrated region was shown in 
Fig. 12. Unlike large scale appearance, the very central region within 300 pc 
could be described evenly spread over the space. Nevertheless, even in this 
closer look the local Galactic disc structure is noticeable when the scaleheight 
lines were drawn on Fig. 12b. One may easily notice that all local binaries 
($d\leq 300$ pc) are contained within the thin-disc scaleheight limit except for 
only three systems.

\subsection{How Good Single Stars were Represented?}
The components of close binary systems with proximity effects are flattened 
at various degrees by fast rotation and tidally elongated towards each other 
by the mutual gravity in addition to irradiation effects. All these effects 
may be negligible in the case of detached binaries, where single star 
evolution could be applied to each component in better approximation. 
It is advantageous to  know how good the present sample represents non-rotating 
spherical single stars of a given mass.  However, the flattening due to rotation, 
the mass loss due to stellar winds may not be ignorable even for the case of 
single stars.  

The tidal evolution forces the components of close binaries into tidal 
synchronization. For interested readers, the catalogue contains projected 
rotational velocities ($v\sin i$), from which one can compute if tidal 
synchronization was achieved. More importantly, rotation and Roche lobe 
filling ratio are two indicators for the component stars deviating from 
sphericity. For the sake of confidence, the sphericity of the present 
sample was inspected by computing the Roche model of each binary.

Mass, mass ratio and orbital period determine the sizes and the shapes of 
the Roche lobes \citep{Kopal78}. Classical definition of close (interacting) 
binary systems was given as $R_1+R_2\geq 0.1a$, where a is the semi-major 
axis of the orbit and $R_{1,2}$ are the radii of the components, the subscripts 
are for the primary and the secondary. With counter definition $R_1+R_2<0.1a$, 
the non-interacting binaries, however, are all well detached systems definitely 
far from  proximity effects and well contained in the Roche lobes that their 
shapes are believed to be spherical. Unfortunately, the number of such systems 
is only 24 in the present sample and much less in previous samples 
\citep[four in the list of][]{Torres10}. In order to increase the number of 
detached systems, one prefers to loosen the condition $R_1+R_2<0.1a$.
 
Equipotential surface of a component star is represented by four fractional 
radii defined as: $r$(side), $r$(back), $r$(point) and $r$(pole). The smallest 
of those is the $r$(pole) and the largest one is the $r$(point). The difference 
between those fractional radii increases by increasing filling ratio. In order to
test and establish a well-structured limit for deviation from the sphericity we 
have computed Roche model of the sample binaries using Binary 
Maker\footnote{http://www.binarymaker.com/}. This allowed us to compare 
fractional radii and made eye inspection of the shapes of the component.

Figure 13 displays how radii and shapes of the
components change with respect to the Roche lobes from 60 per cent filling
to the higher rates of filling factors (FF), which is defined as 
FF=$\bar r/\bar r_{RL}$, where, $\bar r$ is average radius, $\bar r_{RL}$ is 
Roche-lobe radius relative to the semi-major axis. It can be deduced from an 
eye inspection that deviations from sphericity could be ignorable for small
filling factors which can be up to 75 per cent, a value corresponding to
the difference between the $r$(side) and the $r$(pole) being less than
1 per cent. By the way, the difference between the $r$(point) and the 
$r$(side) being larger than 1 per cent indicate tidal elongation and 
tidal syncronization when $P_{rot}=P_{orb}$. By a careful look, eye could 
feel larger deviations (deviation $>1$ per cent corresponding FF $>75$ per 
cent), thus, non-spherical shapes of the component stars of V478 Cyg and 
DH Cep on Fig. 13 are noticeable. On the other hand, with less filling 
factors (FF $<$ 75 per cent), component stars of AG Ari and V760 Sco appear
spherical.

The distribution of the filling ratios of the component stars in the
present sample is displayed in Fig. 14. Accordingly, 
426 stars (82.9 per cent of the sample) have filing factors less than
70 per cent, which could be assumed to have spherical shapes. If one
could expand sphericity limit to 75 per cent filling factor, the number
of spherical stars increases to 455 (88.5 per cent of the sample).
Overall, 20 binaries were found as both components exceed the limit of
75 per cent filling factor, which includes seven O-type binaries. The
number of binaries at least one component exceeding the limit is 39,
which includes nine O-type binaries. That is, among the 39 there are
19 systems only one component is spherical. We first tempted to discard
the systems with non-spherical components. However, the number of O-type binaries
is only 10 (3.9 per cent of the sample). Among the 10, only one (Y Cyg, 
with $P_{orb}=2.996$ days) were found to have both components with spherical 
shapes. Therefore, we have decided not to discard the systems with more 
than 75 per cent filling factor and left this choice to researchers, who 
may alter sphericity limit according to the their specific needs.

\section{Conclusions}

\begin{itemize} 
  \item The most accurate stellar parameters (masses, radii, temperatures, 
   surface gravities, luminosities, projected rotational velocities, radial 
   velocity amplitudes, mass ratio, orbital period, orbital inclination, 
   semi-major axis, eccentricity) were compiled from the simultaneous solutions 
   of light and radial velocity curves of detached double-lined eclipsing binaries. 

 \item The masses and radii were homogenized using most recent values of 
  $GM_{\odot}=1.3271244\times10^{20}$ m$^{3}$s$^{-2}$ \citep{Standish95} and 
  $R_{\odot}=6.9566\times10^{8}$m \citep{Haberreiter08}. 
  Surface gravities and luminosities recomputed using homogenized masses and radii.

 \item Apparent magnitudes, orbital periods, masses, radii, mass ratios, 
       spectral types and space distribution of the present sample were discussed.

 \item The number of stars with both mass and radius as accurate as 1 per cent 
       is 93, as accurate as 3 per cent is 311, and as accurate as 5 per cent 
       is 388.

 \item Filling ratios of the current sample were studied. Thus, 
  the geometrical shapes of the component stars were determined. Up to 75 
  per cent of filling factors, stars are found almost spherical within 
  1 per cent uncertainty.
     
 \item Giants and supergiants are missing in the present sample. 
  Observational astronomers are encouraged to explore eclipsing binaries 
  among giants and supergiants. Improving light curve observing techniques 
  for discovering small amplitude SB2 eclipsing systems is a challenge

\end{itemize} 

\section{Acknowledgments}
This work has been supported in part by the Scientific and Technological 
Research Council of Turkey (T\"UB\.ITAK) grant numbers 106T688 and 111T224. 
Authors would like to thank anonymous referee who provided valuable comments 
for improving the manuscript and Mr. Muzaffer Karasulu for proof reading. 
This research has made use of the SIMBAD database, operated at CDS, Strasbourg, 
France and NASA's Astrophysics Data System. We would like to thank Dr. 
Nilda Oklay for helping online material.

% Table 2
\begin{table*}
\setcounter{table}{1} 
\center 
\caption{Column descriptions of the catalogue data.} 
\begin{tabular}{cll}
\hline
    Column & Description &     Remark \\
\hline
         1 &         ID & Order number \\
         2 &  Star name & The most common name \\
         3 &         HD & Henry Draper catalogue number \\
         4 &        Hip & {\em Hipparcos} catalogue number \\
         5 &        SAO & Smithsonian Astrophysical Observatory star catalogue number \\
         6 & $\alpha$ (hh:mm:ss.ss) & Right ascension in J2000 \\
         7 & $\delta$ (dd:mm:ss.ss) & Declination in J2000 \\
\hline
\end{tabular}  
\end{table*}

% Table 3
\begin{landscape}
\textwidth = 650pt 
\begin{table*}
\setcounter{table}{2} 
\setlength{\tabcolsep}{1pt}
\tiny{
\begin{center}
\caption{Homogenized parameters and orbital sizes of detached, double-lined eclipsing binaries.}
\begin{tabular}{llccccccccccc}
\hline
        ID &     System &   $M_1$ &   $M_2$ &   $R_1$ &   $R_2$ &  $T_1$ &   $T_2$ & $\log L_1$ & $\log L_2$ & $\log g_1$ & $\log g_2$ &    $a$ \\
& &($M/M_{\odot}$) &($M/M_{\odot}$)&($R/R_{\odot}$)&($R/R_{\odot}$)&  (K) &   (K) & ($L/L_{\odot}$) &($L/L_{\odot}$) & (cgs) & (cgs) & ($R_{\odot}$) \\
\hline
         1 &     DV Psc              & 0.677$\pm$0.019 & 0.475$\pm$0.010 & 0.685$\pm$0.030 & 0.514$\pm$0.020 &  4450$\pm$8   & 3614$\pm$8    &-0.782$\pm$0.040 &-1.393$\pm$0.035 & 4.597$\pm$0.042 & 4.693$\pm$0.037 &  2.015$\pm$0.101 \\
         2 &     MU Cas              & 4.657$\pm$0.100 & 4.575$\pm$0.090 & 4.192$\pm$0.050 & 3.671$\pm$0.040 & 14750$\pm$500 & 15100$\pm$500 & 2.873$\pm$0.064 & 2.799$\pm$0.063 & 3.861$\pm$0.014 & 3.969$\pm$0.013 & 40.026$\pm$0.190 \\
         3 & V69-47 Tuc              & 0.876$\pm$0.005 & 0.859$\pm$0.006 & 1.316$\pm$0.005 & 1.163$\pm$0.006 & 5945$\pm$150  & 5959$\pm$150  & 0.288$\pm$0.046 & 0.185$\pm$0.046 & 4.142$\pm$0.004 & 4.241$\pm$0.005 & 48.324$\pm$0.010 \\
         4 &     YZ Cas              & 2.308$\pm$0.010 & 1.347$\pm$0.010 & 2.547$\pm$0.030 & 1.359$\pm$0.020 & 9200$\pm$300  & 6700$\pm$250  & 1.620$\pm$0.062 & 0.524$\pm$0.072 & 3.989$\pm$0.011 & 4.301$\pm$0.013 & 17.584$\pm$0.070 \\
         5 & NGC188 KR V12           & 1.102$\pm$0.007 & 1.080$\pm$0.007 & 1.425$\pm$0.019 & 1.374$\pm$0.019 & 5900$\pm$100  & 5875$\pm$100  & 0.344$\pm$0.033 & 0.305$\pm$0.033 & 4.173$\pm$0.012 & 4.196$\pm$0.013 & 19.021$\pm$0.040 \\
         6 &   V364 Cas              & 1.684$\pm$0.094 & 1.559$\pm$0.095 & 1.600$\pm$0.010 & 1.600$\pm$0.010 & 7816$\pm$86   & 7780$\pm$86   & 0.933$\pm$0.020 & 0.925$\pm$0.020 & 4.256$\pm$0.026 & 4.223$\pm$0.028 &  8.318$\pm$0.010 \\
\hline
\end{tabular}  
\end{center}
}
\end{table*}
\end{landscape}

\begin{landscape}
\textwidth = 650pt 
\begin{table*}
\setcounter{table}{2} 
\setlength{\tabcolsep}{1pt}
%\contcaption
\tiny{
\begin{center}
\caption{cont.}
\begin{tabular}{llccccccccccc}
\hline
        ID &     System &   $M_1$ &   $M_2$ &   $R_1$ &   $R_2$ &  $T_1$ &   $T_2$ & $\log L_1$ & $\log L_2$ & $\log g_1$ & $\log g_2$ &    $a$ \\
& &($M/M_{\odot}$) &($M/M_{\odot}$)&($R/R_{\odot}$)&($R/R_{\odot}$)&  (K) &   (K) & ($L/L_{\odot}$) &($L/L_{\odot}$) & (cgs) & (cgs) & ($R_{\odot}$) \\
\hline
        81 &     CN Lyn               & 1.038$\pm$0.020 & 1.038$\pm$0.020 & 1.799$\pm$0.210 & 1.839$\pm$0.240 & 6500$\pm$250 & 6455$\pm$260 & 0.715$\pm$0.142 & 0.722$\pm$0.159 & 3.944$\pm$0.116 & 3.925$\pm$0.132 & 8.396$\pm$0.050\\
        82 &     AI Hya               & 1.974$\pm$0.040 & 2.140$\pm$0.040 & 2.764$\pm$0.020 & 3.912$\pm$0.030 & 7100$\pm$60  & 6700$\pm$60  & 1.241$\pm$0.016 & 1.442$\pm$0.017 & 3.850$\pm$0.011 & 3.584$\pm$0.011 &27.620$\pm$0.170\\
        83 & NSVS 07394765            & 0.360$\pm$0.005 & 0.180$\pm$0.004 & 0.459$\pm$0.004 & 0.491$\pm$0.005 & 3300$\pm$200 & 3106$\pm$125 &-1.649$\pm$0.121 &-1.696$\pm$0.077 & 4.671$\pm$0.010 & 4.311$\pm$0.013 & 5.912$\pm$0.505\\
        84 & TYC 5998-1918-1          & 0.703$\pm$0.003 & 0.687$\pm$0.003 & 0.694$\pm$0.009 & 0.699$\pm$0.013 & 4350$\pm$200 & 4090$\pm$200 &-0.810$\pm$0.089 &-0.911$\pm$0.096 & 4.602$\pm$0.012 & 4.586$\pm$0.017 & 6.232$\pm$0.010\\
        85 &     AY Cam               & 1.947$\pm$0.041 & 1.707$\pm$0.036 & 2.784$\pm$0.020 & 2.034$\pm$0.015 & 7250$\pm$100 & 7395$\pm$100 & 1.284$\pm$0.025 & 1.046$\pm$0.025 & 3.838$\pm$0.011 & 4.054$\pm$0.011 &12.677$\pm$0.090\\
        86 &     VV Pyx               & 2.097$\pm$0.018 & 2.094$\pm$0.018 & 2.169$\pm$0.020 & 2.169$\pm$0.200 & 9500$\pm$200 & 9500$\pm$200 & 1.537$\pm$0.039 & 1.537$\pm$0.098 & 4.087$\pm$0.009 & 4.087$\pm$0.089 &18.757$\pm$0.050\\
\hline
\end{tabular}
\end{center}
}
\end{table*}
\end{landscape}

\begin{landscape}
\textwidth = 650pt 
\begin{table*}
\setcounter{table}{2} 
\setlength{\tabcolsep}{1pt}
%\contcaption
\tiny{
\begin{center}
\caption{cont.}
\begin{tabular}{llccccccccccc}
\hline
        ID &     System &   $M_1$ &   $M_2$ &   $R_1$ &   $R_2$ &  $T_1$ &   $T_2$ & $\log L_1$ & $\log L_2$ & $\log g_1$ & $\log g_2$ &    $a$ \\
& &($M/M_{\odot}$) &($M/M_{\odot}$)&($R/R_{\odot}$)&($R/R_{\odot}$)&  (K) &   (K) & ($L/L_{\odot}$) &($L/L_{\odot}$) & (cgs) & (cgs) & ($R_{\odot}$) \\
\hline
       161 &  V1647 Sgr               & 2.184$\pm$0.040 & 1.967$\pm$0.030 & 1.908$\pm$0.021 & 1.741$\pm$0.021 & 9600$\pm$310  & 9100$\pm$300  & 1.443$\pm$0.061 & 1.271$\pm$0.063 & 4.216$\pm$0.013 & 4.250$\pm$0.013 & 14.940$\pm$0.083 \\
       162 &  V3903 Sgr               &27.204$\pm$0.549 &18.964$\pm$0.439 & 8.120$\pm$0.086 & 6.149$\pm$0.060 &38000$\pm$1900 &34100$\pm$1700 & 5.091$\pm$0.098 & 4.662$\pm$0.097 & 4.054$\pm$0.013 & 4.138$\pm$0.013 & 21.876$\pm$0.211 \\
       163 &     EG Ser               & 2.200$\pm$0.050 & 1.990$\pm$0.030 & 1.689$\pm$0.010 & 1.549$\pm$0.010 & 9900$\pm$200  & 9100$\pm$200  & 1.391$\pm$0.037 & 1.169$\pm$0.040 & 4.325$\pm$0.011 & 4.357$\pm$0.009 & 31.383$\pm$0.200 \\
       164 & V994 Her-B               & 1.868$\pm$0.120 & 1.859$\pm$0.120 & 1.583$\pm$0.080 & 1.493$\pm$0.080 & 9000$\pm$250  & 8450$\pm$70   & 1.169$\pm$0.071 & 1.009$\pm$0.052 & 4.311$\pm$0.055 & 4.359$\pm$0.058 &  8.243$\pm$0.149 \\
       165 & V994 Her-A               & 2.828$\pm$0.200 & 2.299$\pm$0.160 & 2.123$\pm$0.049 & 1.688$\pm$0.039 &12000$\pm$250  & 9450$\pm$90   & 1.924$\pm$0.043 & 1.310$\pm$0.027 & 4.236$\pm$0.038 & 4.345$\pm$0.038 & 11.837$\pm$0.247 \\
       166 &   V451 Oph               & 2.769$\pm$0.060 & 2.351$\pm$0.050 & 2.646$\pm$0.030 & 2.032$\pm$0.030 &10800$\pm$800  & 9800$\pm$500  & 1.932$\pm$0.153 & 1.534$\pm$0.100 & 4.035$\pm$0.014 & 4.194$\pm$0.016 & 12.257$\pm$0.100 \\
\hline
\end{tabular}
\end{center}
}
\end{table*}
\end{landscape}

\begin{landscape}
\textwidth = 650pt 
\begin{table*}
\setcounter{table}{2} 
\setlength{\tabcolsep}{1pt}
%\contcaption
\tiny{
\begin{center}
\caption{cont.}
\begin{tabular}{llccccccccccc}
\hline
        ID &     System &   $M_1$ &   $M_2$ &   $R_1$ &   $R_2$ &  $T_1$ &   $T_2$ & $\log L_1$ & $\log L_2$ & $\log g_1$ & $\log g_2$ &    $a$ \\
& &($M/M_{\odot}$) &($M/M_{\odot}$)&($R/R_{\odot}$)&($R/R_{\odot}$)&  (K) &   (K) & ($L/L_{\odot}$) &($L/L_{\odot}$) & (cgs) & (cgs) & ($R_{\odot}$) \\
\hline
       241 &     AH Cep              &15.565$\pm$0.202 &13.714$\pm$0.202 & 6.385$\pm$0.110 & 5.864$\pm$0.130 & 29900$\pm$1000& 28600$\pm$1000& 4.466$\pm$0.065 & 4.315$\pm$0.069 & 4.020$\pm$0.016 & 4.039$\pm$0.021 & 19.014$\pm$0.140 \\
       242 &   V364 Lac              & 2.334$\pm$0.015 & 2.296$\pm$0.025 & 3.308$\pm$0.038 & 2.986$\pm$0.035 &  8250$\pm$150 &  8500$\pm$150 & 1.658$\pm$0.034 & 1.621$\pm$0.034 & 3.767$\pm$0.010 & 3.849$\pm$0.011 & 26.519$\pm$0.080 \\
       243 &   V453 Cep              & 2.575$\pm$0.050 & 2.478$\pm$0.050 & 2.112$\pm$0.201 & 2.011$\pm$0.201 & 10300$\pm$500 & 10400$\pm$500 & 1.654$\pm$0.138 & 1.628$\pm$0.141 & 4.200$\pm$0.092 & 4.225$\pm$0.097 &  8.086$\pm$0.060 \\
       244 &     NY Cep              &13.152$\pm$1.012 & 9.323$\pm$1.036 & 6.843$\pm$0.704 & 5.434$\pm$0.503 & 28500$\pm$1000& 23100$\pm$1000& 4.443$\pm$0.124 & 3.878$\pm$0.127 & 3.887$\pm$0.108 & 3.937$\pm$0.106 & 73.119$\pm$2.435 \\
       245 &     EF Aqr              & 1.244$\pm$0.008 & 0.946$\pm$0.006 & 1.338$\pm$0.012 & 0.956$\pm$0.012 &  6150$\pm$65  &  5185$\pm$110 & 0.362$\pm$0.020 &-0.227$\pm$0.040 & 4.280$\pm$0.008 & 4.453$\pm$0.011 & 10.996$\pm$0.020 \\
       246 &     CW Cep              &11.797$\pm$0.140 &11.067$\pm$0.140 & 5.524$\pm$0.121 & 5.030$\pm$0.121 & 28300$\pm$1000& 27700$\pm$1000& 4.245$\pm$0.070 & 4.126$\pm$0.072 & 4.025$\pm$0.020 & 4.079$\pm$0.022 & 23.326$\pm$0.081 \\
\hline
\end{tabular}
\end{center}
}
\end{table*}
\end{landscape}

% Table 4
\begin{table*}
\setcounter{table}{3} 
\setlength{\tabcolsep}{5pt}
\begin{center}
\caption{Summary statistics of the present sample.}
\begin{tabular}{cccccccc}
\hline
   &       Mass &     Radius &    $T_{eff}$ &   $\log g$ &          Luminosity &     $a$ &    $q$    \\
   & ($M/M_{\odot}$) & ($R/R_{\odot}$) &   (K) &(cms$^{-2}$)& ($L/L_{\odot}$) &     ($R_\odot$) &    \\
\hline
   Maximum &      32.60 &      21.19 &      43000 &      5.077 &      5.554 &     233.25 &     1.114 \\
      Star (or system)  &        (1) &        (2) &       (3)  &       (4)  &        (1) &        (5) &      (6)  \\
   Minimum &       0.18 &       0.21 &       2750 &      2.906 &     -2.553 &       1.85 &     0.237 \\
      Star (or system)  &        (7) &        (8) &        (9) &      (10)  &        (8) &       (11) &     (12)  \\
      Mean &        2.9 &        2.1 &       9500 &       4.20 &        2.8 &       18.9 &     0.851 \\
      Mode &        1.3 &        1.3 &       6500 &       4.25 &        4.4 &       10.5 &     0.975 \\
    Median &        1.5 &        1.7 &       6800 &       4.21 &        3.9 &       14.1 &     0.905 \\
\hline
\multicolumn{8}{c}{Errors}\\
%    Errors &        &        &         &        &       &        &      \\
\hline  
   Minimum &    0.09\% &   0.19\% &          6 &      0.003 &       0.5\% &    0.01\% &     0.001 \\
   Maximum &    16.8\% &   51.9\% &       4100 &      1.934 &        51\% &    8.54\% &     0.178 \\
      Mean &     2.4\% &    3.2\% &        282 &      0.044 &      12.5\% &    0.44\% &     0.001 \\
      Mode &     1.1\% &    0.9\% &        125 &      0.010 &         9\% &    0.02\% &     0.004 \\
    Median &     1.3\% &    1.5\% &        150 &      0.016 &      10.1\% &    0.86\% &     0.007 \\
\hline
\end{tabular} 
\end{center}
(1) Primary of DH Cep; (2) primary of V1765 Cyg; (3) primary of V1182 Aql; (4) secondary of 2MASS J04463285+1901432;
(5) V379 Cep; (6) V885 Cyg; (7) secondary of NSVS 07394765;  (8) secondary of 2MASS J04463285+1901432; 
(9) secondary of KIC 10935310; (10) secondary of TZ For; (11) SDSS-MEB-1; (12) TYC 3121-1659-1.\\
\end{table*}

% Table 5
\begin{table*}
\setcounter{table}{4} 
\setlength{\tabcolsep}{5pt}
\begin{center}
\caption{Distribution of orbital periods among the spectral types of the primary components.}
\begin{tabular}{ccccccccc}
\hline
Spectral Type         &   O &    B &    A &    F &    G &    K &    M & Total \\
\hline
$-0.6\leq\log P_{orb}<-0.2$ &     &      &      &    1 &    3 &    5 &    5 &    14 \\
$-0.2\leq\log P_{orb}<0.2$  &     &    5 &    3 &    9 &    7 &    6 &    7 &    37 \\
$0.2\leq\log P_{orb}<0.6$   &   9 &   28 &   28 &   21 &    6 &    6 &    7 &   105 \\
$0.6\leq\log P_{orb}<1.0$   &   1 &   11 &   16 &   25 &   12 &    3 &    1 &    69 \\
$1.0\leq\log P_{orb}<1.4$   &     &    7 &    7 &    5 &    5 &      &      &    24 \\
$1.4\leq\log P_{orb}<1.8$   &     &      &    3 &    1 &    1 &      &    1 &     6 \\
$1.8\leq\log P_{orb}<2.2$   &     &    1 &      &    1 &      &      &      &     2 \\
\hline
    Total             &  10 &   52 &   57 &   63 &   34 &   20 &   21 &   257 \\
\hline
    Median $\log P_{orb}$   &0.44 & 0.43 & 0.53 & 0.60 & 0.67 & 0.18 & 0.18 &       \\
Median $q$    &0.708& 0.830& 0.909& 0.931& 0.922& 0.926& 0.922&  \\
\hline
\end{tabular}
\end{center}
\end{table*}

%FIGURE 1
\begin{figure*}
\begin{center}
\includegraphics[scale=0.60, angle=0]{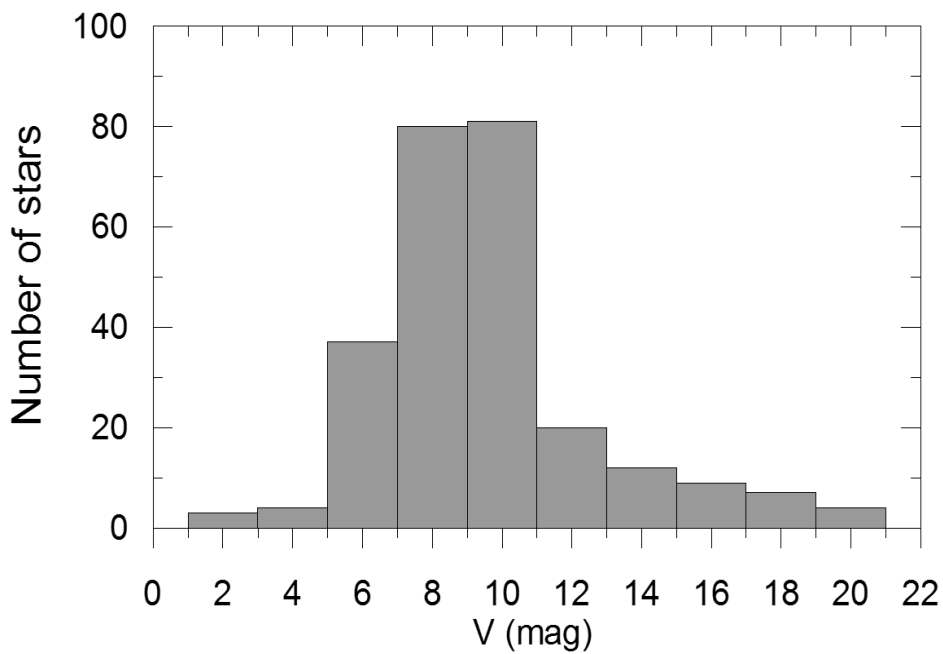}
\caption[] {Apparent brightness distribution of 257 systems.}
\end{center}
\end{figure*}

%FIGURE 2
\begin{figure*}
\begin{center}
\includegraphics[scale=0.60, angle=0]{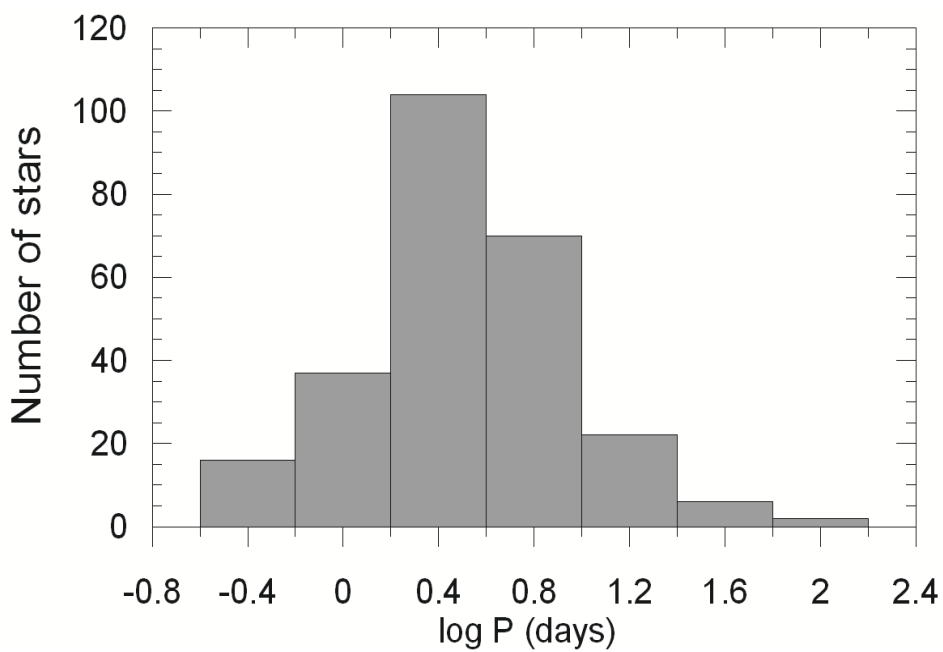}
\caption[] {Orbital period distribution of 257 systems.}
\end{center}
\end{figure*}

%FIGURE 3
\begin{figure*}
\begin{center}
\includegraphics[scale=0.50, angle=0]{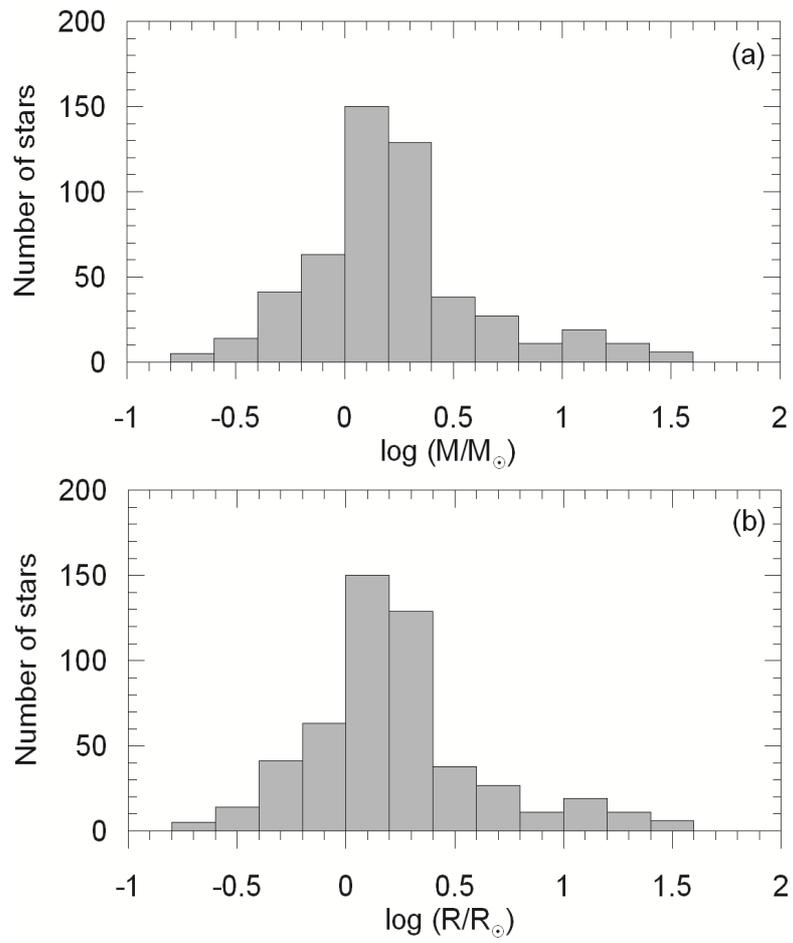}
\caption[] {Distributions of masses (a) and radii (b) of 514 stars from 257 binaries.}
\end{center}
\end{figure*}

%FIGURE 4
\begin{figure*}
\begin{center}
\includegraphics[scale=0.80, angle=0]{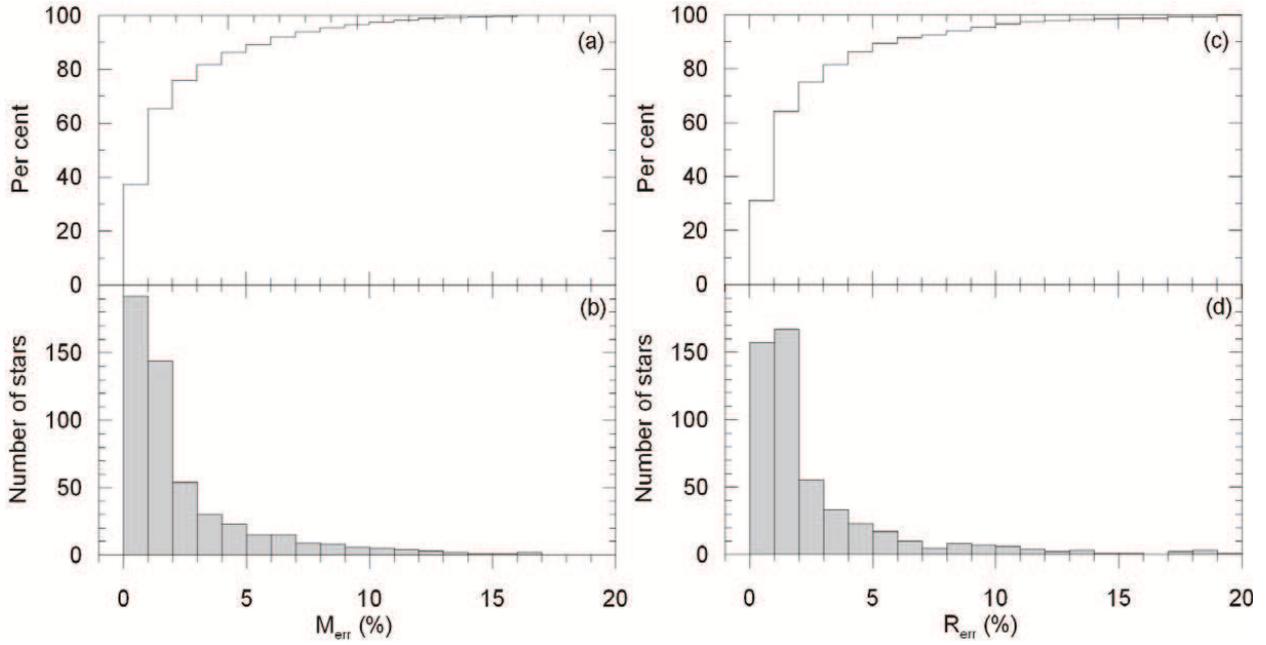}
\caption[] {Uncertainty distributions of masses (a, b) and radii (c, d).}
\end{center}
\end{figure*}

%FIGURE 5
\begin{figure*}
\begin{center}
\includegraphics[scale=0.60, angle=0]{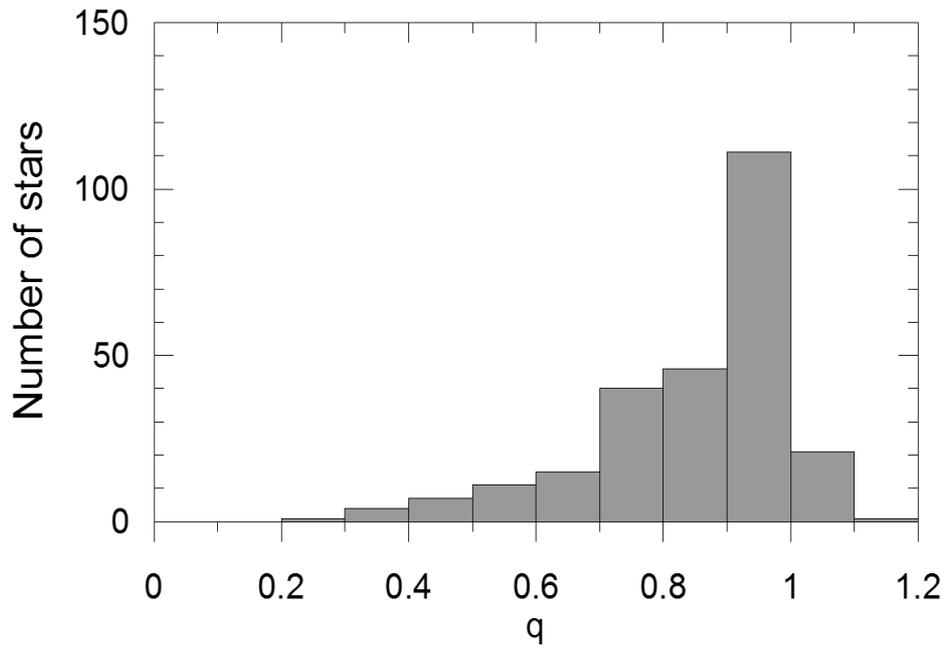}
\caption[] {Mass ratio distribution of the sample. The median of $q$ is 0.905.}
\end{center}
\end{figure*}

%FIGURE 6
\begin{figure*}
\begin{center}
\includegraphics[scale=0.40, angle=0]{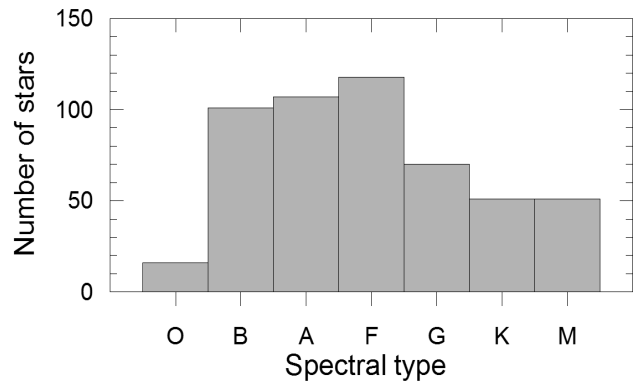}
\caption[] {Distribution of 514 stars (257 pairs) according to spectral types.}
\end{center}
\end{figure*}

%FIGURE 7
\begin{figure*}
\begin{center}
\includegraphics[scale=0.55, angle=0]{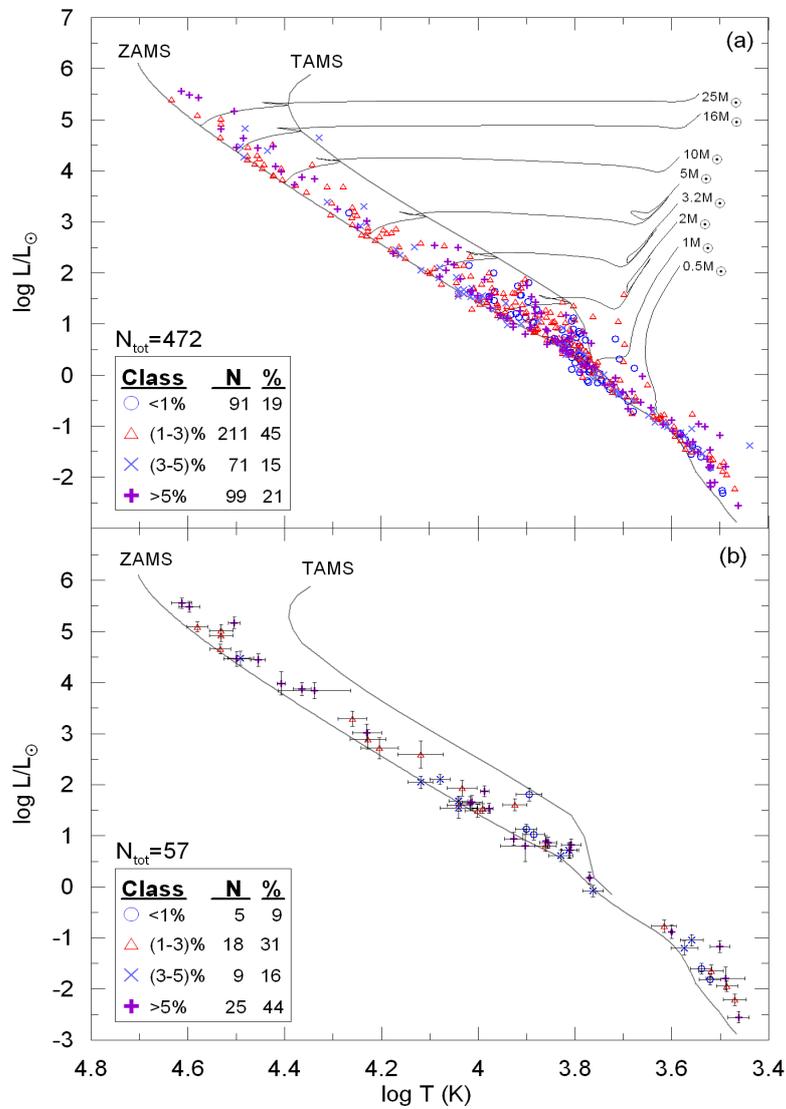}
\caption[] {H-R diagram of 472 stars (a) and error bars (b) of 57 least accurate stars of the sample. Displayed error bars imply that most of the error bars in (a) are smaller than the symbols printed.}
\end{center}
\end{figure*}

%FIGURE 8
\begin{figure*}
\begin{center}
\includegraphics[scale=0.40, angle=0]{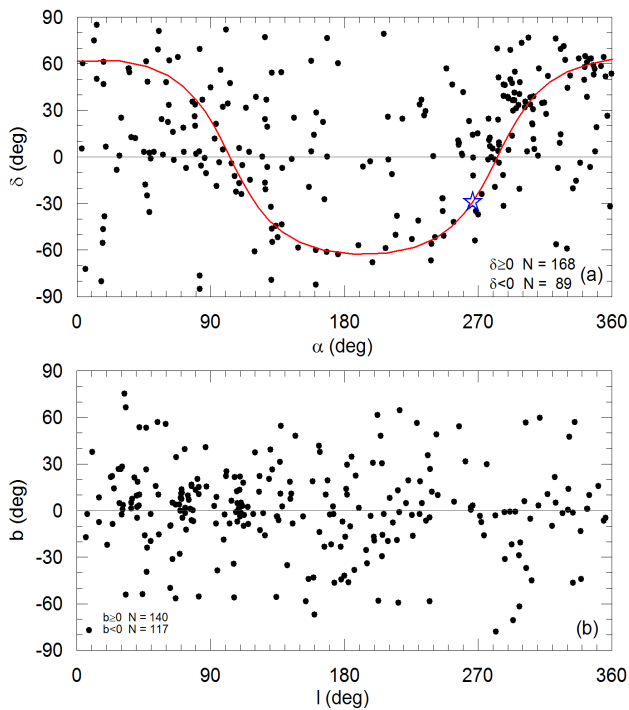}
\caption[] {Distributions on the equatorial (a) and the Galactic (b) coordinates. 
Sinusoidal line in the upper panel is the Galactic plane, where the Galactic 
center is marked as a big star symbol.}
\end{center}
\end{figure*}

%FIGURE 9
\begin{figure*}
\begin{center}
\includegraphics[scale=0.40, angle=0]{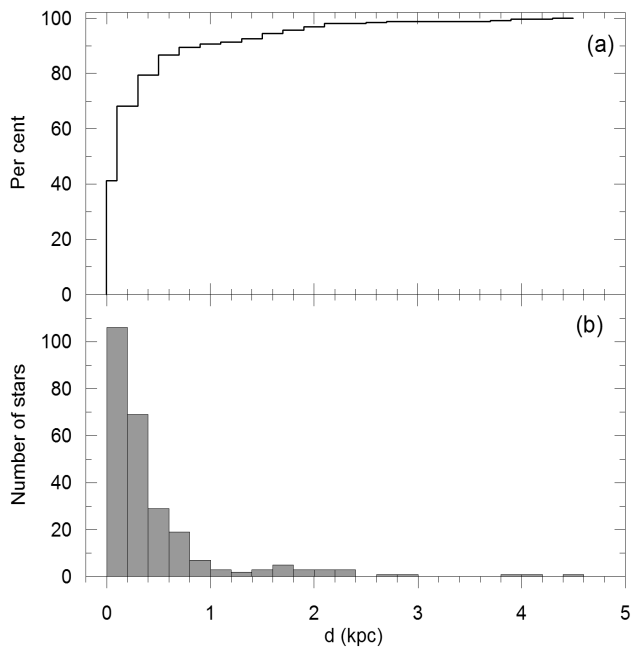}
\caption[] {Distance distributions of the sample stars in the Solar 
neighborhood. (a) is in  cumulative and (b) is in frequencies.}
\end{center}
\end{figure*}

%FIGURE 10
\begin{figure*}
\begin{center}
\includegraphics[scale=0.40, angle=0]{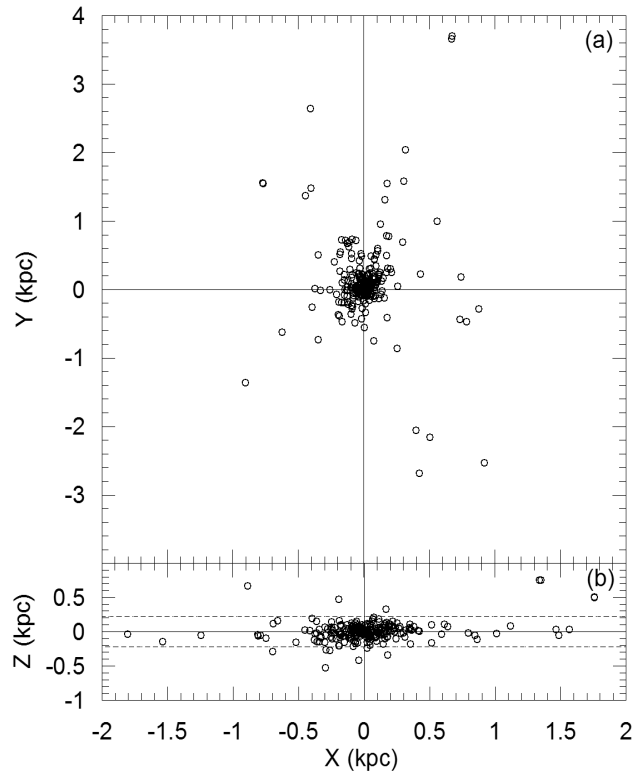}
\caption[] {Distribution on the Galactic plane (a). $X$ in the direction 
towards the Galactic center, $Y$ in the Galactic rotation. Distribution 
on the plane ($X$-$Z$) perpendicular to the Galactic disc (b). 
Dashed line represents the scaleheight of the thin disc. The scaleheight 
of the thin disc ($H=220$ pc) is taken from \citet{Bilir06a, Bilir06b}.}
\end{center}
\end{figure*}

%FIGURE 11
\begin{figure*}
\begin{center}
\includegraphics[scale=0.40, angle=0]{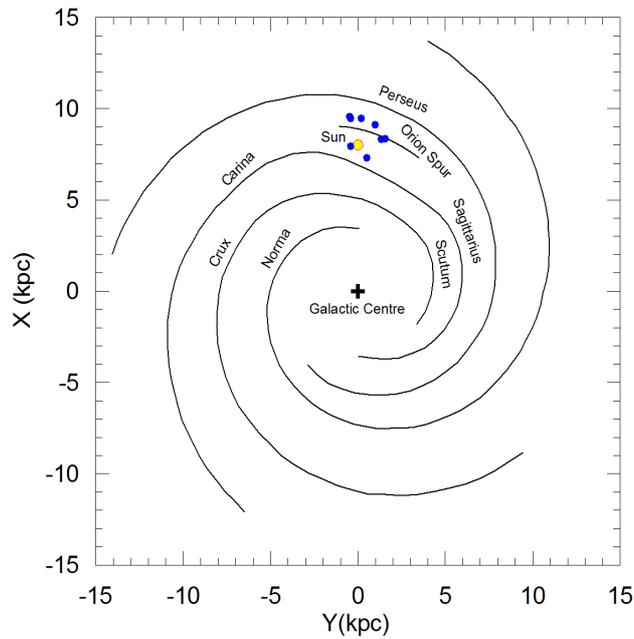}
\caption[] {O-type binaries on the Galactic plane. The position of the spiral 
arms taken from \citet{Xu09}. It is assumed that the distance to the Galactic 
centre of the Sun is 8 kpc.}
\end{center}
\end{figure*}

%FIGURE 12
\begin{figure*}
\begin{center}
\includegraphics[scale=0.70, angle=0]{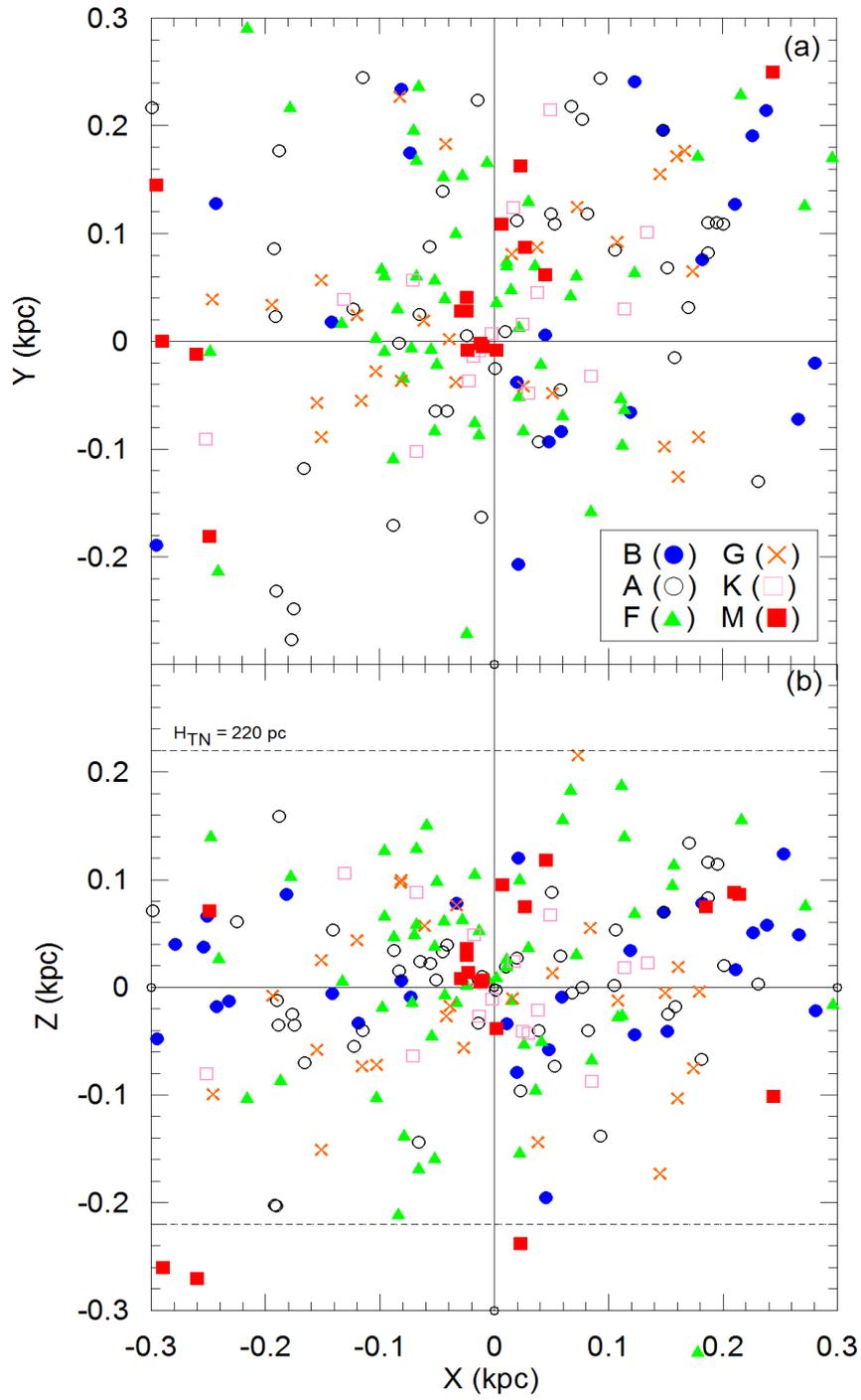}
\caption[] {Nearby distributions of sample binaries within 300 pc.}
\end{center}
\end{figure*}

%FIGURE 13
\begin{figure*}
\begin{center}
\includegraphics[scale=0.45, angle=0]{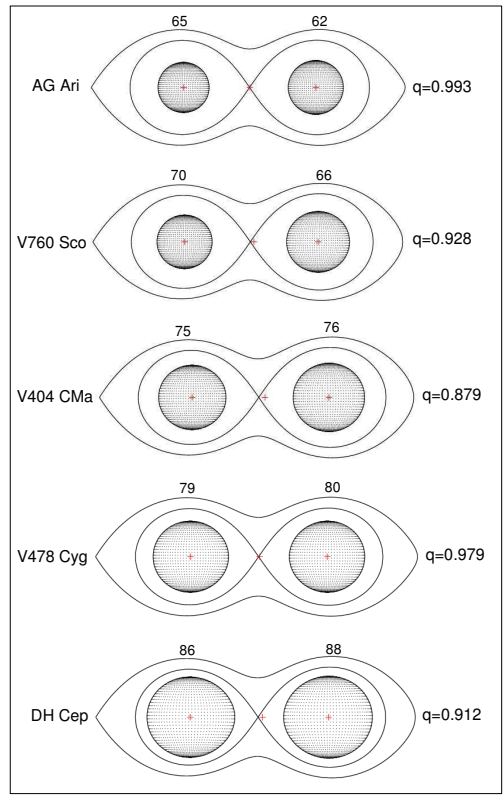}
\caption[] {Examples of Roche lobe filling ratios. Names and the mass ratios are at 
the sides, while filling ratio (FF, Filling Factor) written on the top. By an eye 
inspection, one can notice sphericity spoils after filling factor 75.}
\end{center}
\end{figure*}

%FIGURE 14
\begin{figure*}
\begin{center}
\includegraphics[scale=0.50, angle=0]{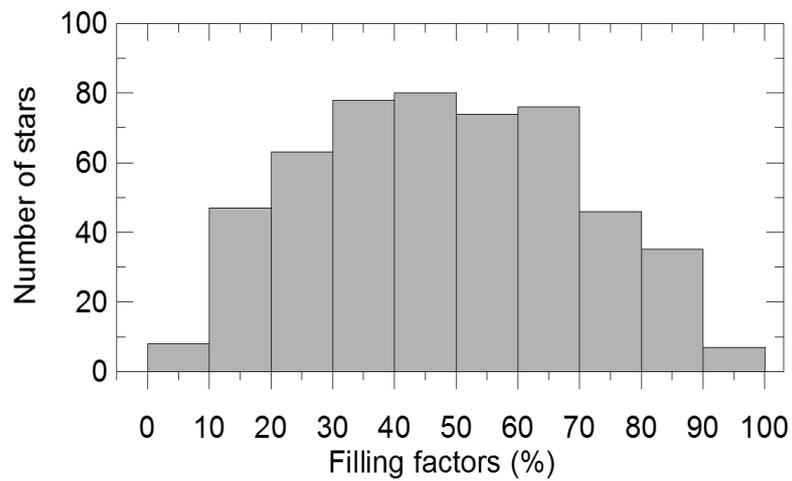}
\caption[] {Distribution of the filling factors among 514 stars (257 binaries) 
of the sample.}
\end{center}
\end{figure*}

\end{document}